\documentclass[fleqn,10pt]{wlscirep}
\usepackage[utf8]{inputenc}
\usepackage[T1]{fontenc}
\usepackage{bm}
\title{Modeling High Entropy Alloys' Mechanical Property through Natural Language-Derived Descriptors}

\author[1]{Li-Cheng Hsiao}
\author[1]{Zi-Kui Liu}
\author[1,2,*]{Wesley Reinhart}

\affil[1]{Materials Sciences and Engineering, Pennsylvania State University}
\affil[2]{Institute for Computational and Data Sciences, Pennsylvania State University}

\affil[*]{e-mail: reinhart@psu.edu}

\begin{abstract}
Processing treatments of alloys, despite being influential to alloy properties, are often neglected in machine-learning aided alloy designs due to the difficulties in expressing these information. We investigated the expressiveness of transformer embedding through synthesized annealing processing treatment text and verified that embeddings could be utilized to reconstruct the processing parameters of alloys effectively with an $R^2$>0.99. We then utilized the vector representations of alloys' processing treatment description as descriptors to model high-entropy alloys' hardness and achieved a ~20\% improvement in prediction, verifying that natural language-derived descriptors of processing treatment information could be utilized to improve prediction of alloy properties.

\end{abstract}

\usepackage{wrapfig}
\usepackage{float}
\begin{document}

\flushbottom
\maketitle

\thispagestyle{empty}

\section*{Introduction}

High-entropy alloys (HEA) are metallic systems formed by mixing multiple near-equimolar elements, which produces unusually high configurational entropy and stabilizes simple solid-solution phases \cite{yeh2004nanostructured, cantor2004microstructural}. Their chemical and structural complexity gives rise to distinctive effects — sluggish diffusion, severe lattice distortion, and synergistic strengthening (“cocktail effect”) \cite{ye2016high} — enabling microstructures and mechanical properties often superior to conventional alloys. These features, combined with broad compositional freedom, make HEA promising frontiers in alloy design and prospective structural materials \cite{yeh2013alloy}.

Despite these potentials in application, systematic HEA development faces serious challenges. The immense compositional space and complex chemistry–microstructure–property relationships hinder empirical discovery. Conventional computational approaches such as CALPHAD \cite{chen2018database, li2023calphad, yang2020revisit} and DFT\cite{ikeda2019ab, zaddach2013mechanical} have advanced the field but remain constrained: CALPHAD primarily predicts equilibrium phases and struggles with metastable structures, while DFT is computationally intensive and limited to small systems. These constraints pose limitations on the application for computationally driven design and engineering of high entropy alloys.

The development of statistical approaches and the vast amount of experimental and computational data available have spurred interest in the application of machine learning (ML) application in HEA designing to navigate through their broad compositional space\cite{wen2019machine, huang2019machine, rao2022machine, liu2023machine}. Early studies applied classifiers such as neural networks, support vector machines, and genetic algorithms to predict HEA phase stability with promising precision \cite{huang2019machine, li2019machine, islam2018machine}. More recent work integrates active learning \cite{sulley2024accelerating, rao2022machine, li2022towards}, where iterative feedback between modeling and experiment accelerates the discovery of novel compositions with enhanced mechanical properties and enables machine learning to be applied not only for material property prediction but also for direct alloy design. Together, these efforts demonstrate the potential of ML to navigate HEA design spaces much more efficiently than traditional methods.

However, most ML research to date has emphasized chemistry–property relationships, often treating composition as the primary descriptor. In practice, the processing history of HEAs—including thermal treatments, cooling rates, and deformation pathways—profoundly influences the microstructural evolution and phase formation. Unlike composition, processing conditions are sequential and complicated due to the vast amount of processes available for HEA treatment and the parameters that could be tuned in HEA processing treatment. Therefore, it is difficult to encode processing treatment information in conventional tabular datasets. As a result, the processing-property relationships of HEA remain underexplored by current ML frameworks, despite their importance to control alloy performance.

Recently, transformer-based deep learning models have attracted attention in applied sciences\cite{wang2023scientific} for their ability to capture sequential and long-ranged dependencies\cite{vaswani2017attention}. Using self-attention mechanisms, these models can represent ordered inputs more flexibly than traditional ML architectures\cite{lei2024materials}. For example, transformer embeddings trained on scientific text have successfully captured meaningful materials–property correlations beyond explicit human labeling\cite{tshitoyan2019unsupervised} , be applied in inverse design in organic chemistry\cite{jablonka2024leveraging} and biomedical experiment planning\cite{gao2024empowering}. These advances suggest that embeddings from transformer-based models may offer a powerful route to incorporating information into HEA design frameworks.

In this research, the feasibility of utilizing transformer-based embeddings to encode HEA processing information into ML models through text describing HEAs' processing information will be investigated. We consider that in order for HEA processing information to be successfully encoded into ML models the text embeddings must satisfy the following criteria:
\begin{enumerate}
  \item The embeddings are capable of capturing the semantic information related to HEAs' processing parameters in a simple and systematic manner:

  While transformer embeddings are vector representation of text, the text itself contains different aspects of semantic information and it is crucial to ensure that the information related to alloys' processing information that are truly relevant to processing parameters that affects HEAs' physical properties can be captured succesfully.
  \item The embeddings should demonstrate phrasing invariance:

  As text embeddings are utilized as the medium representing the processing information of alloys, they will naturally be influenced by the verbal expression of the said information. As a result, investigating the influence of verbal expression on the ability of embedding at conveying HEAs' processing information will be important to ensure that the embeddings can capture the processing information instead of merely focusing on the verbal expression of information.
  
\end{enumerate}

The investigation of the validity of the embedding vectors as machine learning model features will be investigated in the experiment section through text synthesis and modeling of HEAs' hardness through data gathered from existing peer-reviewed publications

\section*{Experiments}

\subsection*{Text Synthesis Experiment}

\begin{wrapfigure}{r}{0.45\textwidth}
    \centering
    \includegraphics[width=0.43\textwidth]{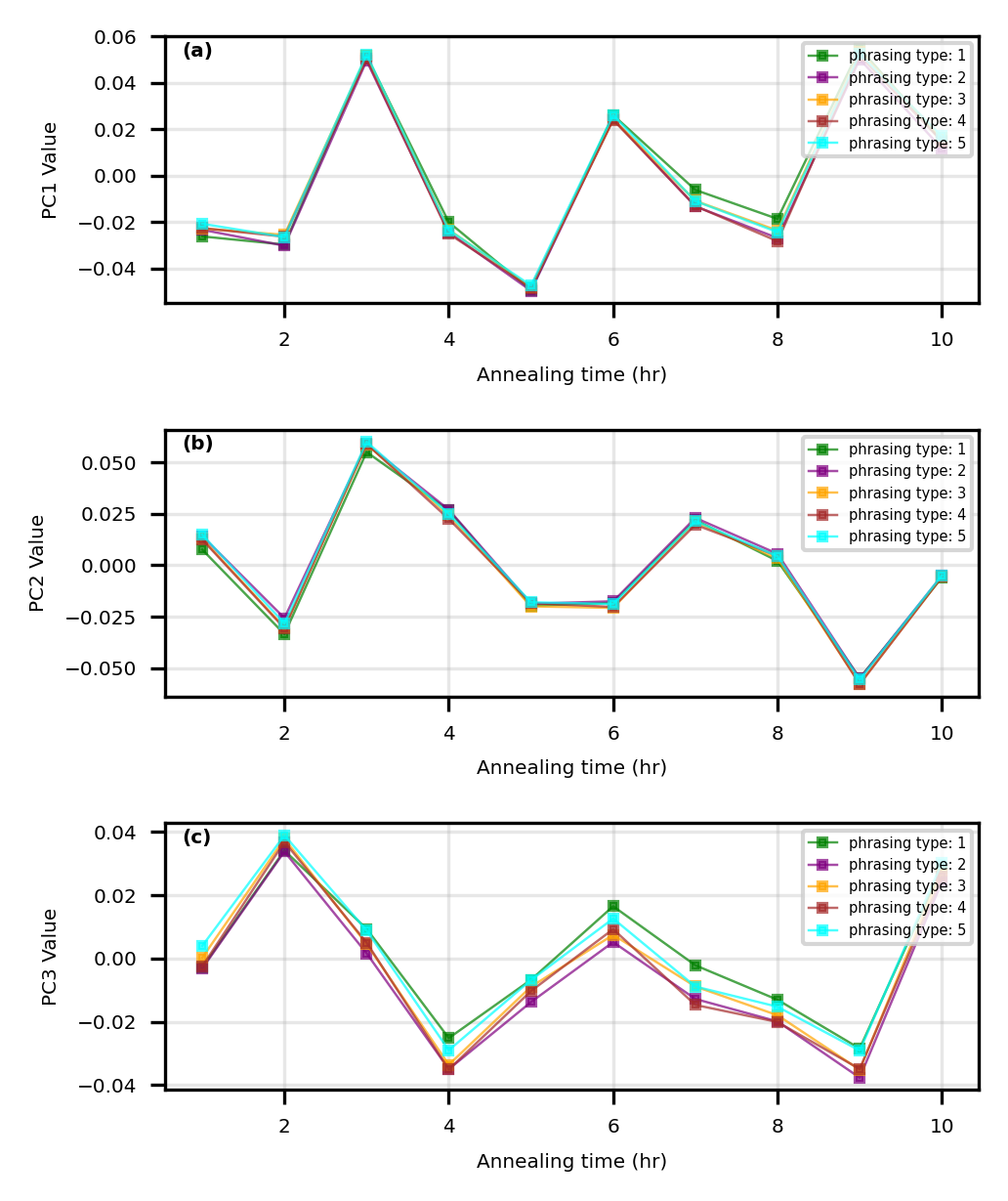}
    \caption{The first (a), the second (b) and the third (c) principal component values of the embeddings of synthesized text with respect to annealing time for annealing description text with different phrasings}
    \label{fig:example}
\end{wrapfigure}

In this section, we will demonstrate that the embeddings are robust and effective features to incorporate HEAs' processing information into machine learning models. That is, the embedding features should be phrasing invariant, representative, and have information well-distributed among features with minimal redundencies. This will be done through the analysis of synthesized text regarding HEAs' annealing treatments.

Text synthesis is conducted to investigate the phrasing invariance and clarity of representation of embedding vectors. In this experiment, text description of HEA processing history is generated through python scripts. The text is aimed to express the semantic meaning of “annealing the alloy under {x} Kelvins and 1 bar for {y} hours”, with x varying from 973 to 1423 in an interval of 50 Kelvins and y varying from 1 to 10. Text with each pair of x and y values are generated with ten different phrasings, totaling an amount of 1000 different text pieces. The text is then transformed into 768-dimensional embeddings through Google Gemini embedding models\cite{team2023gemini} for the assessment of the relationship between embeddings' principal component values and the semantic meaning of their corresponding text.

The values of embeddings' principal components with respect to phrasing types is potted in Fig. 1. It can be observed that despite of the differences in phrasing types in different pieces of text, the values of embeddings' principal components remain similar. Furthermore, the embeddings' principal components demonstrates almost the same trend of change in embeddings' principal components' values. This suggests that while the change in phrasings of the verbal expression of processing information can slightly alter the principal component values, a certain degree of invariance regarding of its semantic information can be successfully retained in their embeddings. 

To further analyze the reconstruction of alloys' processing treatment information through embeddings, linear regression is performed to assess the mapping between semantic information of alloys' processing treatment and embeddings' principal component values. Linear regression models are trained to predict the annealing time and annealing temperature using the principal components of the embeddings of the corresponding text. The regression models demonstrated superb performance, both achieving $\mathrm{R}^2$ greater than 0.99. This suggests that the representation of HEAs' processing treatment information is simple enough to be captured by linear regression models, highlighting the straightforward nature between embedding vectors and the underlying semantic information inherent to their natural language expressions.

\begin{figure}[H]
    \centering
    \includegraphics[width=\textwidth]{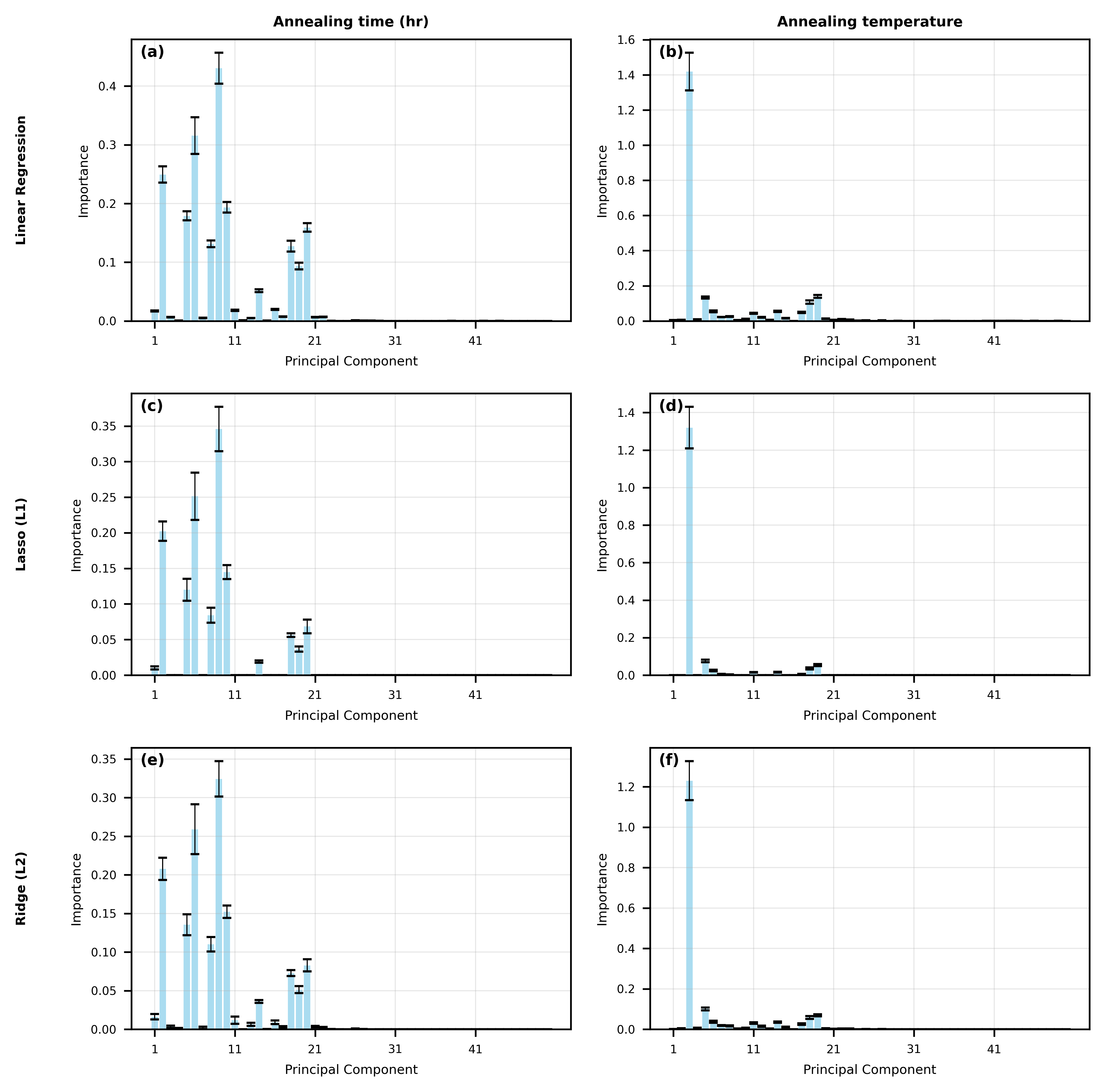}
    \caption{Feature importance analysis of annealing time (a, c, e) and temperature (b, d, f) models under different regularization schemes: Linear (a, b), $\mathrm{L_1}$ (c, d) and $\mathrm{L_2}$ (e, f)}
    \label{fig:fullwidth}
\end{figure}

Feature importance test is conducted to investigate the representation of HEAs' processing treatment information through their embedding vectors. The permutation importance of mundane linear regression models and linear regression models with regularization $\mathrm{L_1}$ and $\mathrm{L_2}$ to predict annealing parameters through the incorporation of principal components has been calculated for both the annealing time and the annealing temperature. Although both annealing time and annealing temperature information can be captured by different dimensions of embedding principal components, annealing temperature requires dimensions with more subtle variances to capture than annealing time, as shown in Fig. 2. As a result, these two different parameters require different amount of principal components to be represented sufficiently well, demonstrating the unequal difficulty of data representation differences in HEA processing treatment parameters. $\mathrm{L_1}$ and $\mathrm{L_2}$ regularization schemes imposed different effects on the feature importances of the linear models, with $\mathrm{L_1}$ regularization introducing feature sparsity while features with minimal importances are retained in $\mathrm{L_2}$-regularized models. However, both regularized models demonstrated similar relative levels of feature importance for almost all important features, which is near identical to un-regularized models. This suggests a distributed representation where multiple latent dimensions jointly contribute to the prediction.

To summarize, the contextual embeddings of the description of HEAs' processing history demonstrate outstanding phrasing invariability, are sufficiently expressive of the processing parameters. The response of the linear models constructed through embeddings also demonstrates the semantic information is neither concentrated in a small number of dimensions nor driven by collinear redundancy and exhibits a distributed and robust encoding instead. All of these behaviors indicate that the contextual embeddings have the potential to serve as a high-quality and robust representation of semantic information of the processing history that supports stable and efficient encoding.

\subsection*{Modeling of HEA Hardness}
In this section, the general effectiveness of utilizing contextual embedding of alloys' processing history will be discussed. The selected dataset is a subset of ULTERA dataset\cite{krajewski2023ultrahigh}, a dataset consisting of over 20,000 datapoints of alloy property data collected from public access publications. A subset of hardness data from ULTERA dataset is chosen to benchmark of the effectiveness of using the contextual embeddings of the description of HEA samples' processing treatments, the datapoints in the dataset of this work are experimentally acquired alloy samples with distinct composition and processing treatment method with their hardness measured at specific temperatures. To ensure that the modeling using this dataset can successfully reflect the effectiveness of contextual embeddings in representing alloy processing treatment information, all of the selected samples consists of at least four constituent elements. In order to address the systematic bias attributed to the differences in experimental methods among different publications, no more than 20 data points could be selected from any particular publication.

The datapoints then go through a final filtering based on their generalized processing treatment types. In ULTERA dataset, the datapoints are labeled with one-hot encodings according to their processing treatment acronyms. For example, an alloy with value on "AC" equal to one indicate that the alloy is "as-cast" without any further treatments, while an alloy with column "HT" and "A" equal to one and zero on all the other processing columns indicate that the alloy sample underwent heat treatment and annealing processes. In the final step of data filtering, only alloys with particular processing treatments are included. The distribution of alloy processing treatment types and constituent elements of the curated dataset are plotted in Fig. 3. For each datapoint (alloy sample), their processing treatment information described in the respective publications they are originally from are extracted using Gemini 2.5 flash models into text and transformed into their corresponding contextual embeddings for further modeling.

\begin{figure}[H]
    \centering
    \includegraphics[width=\textwidth]{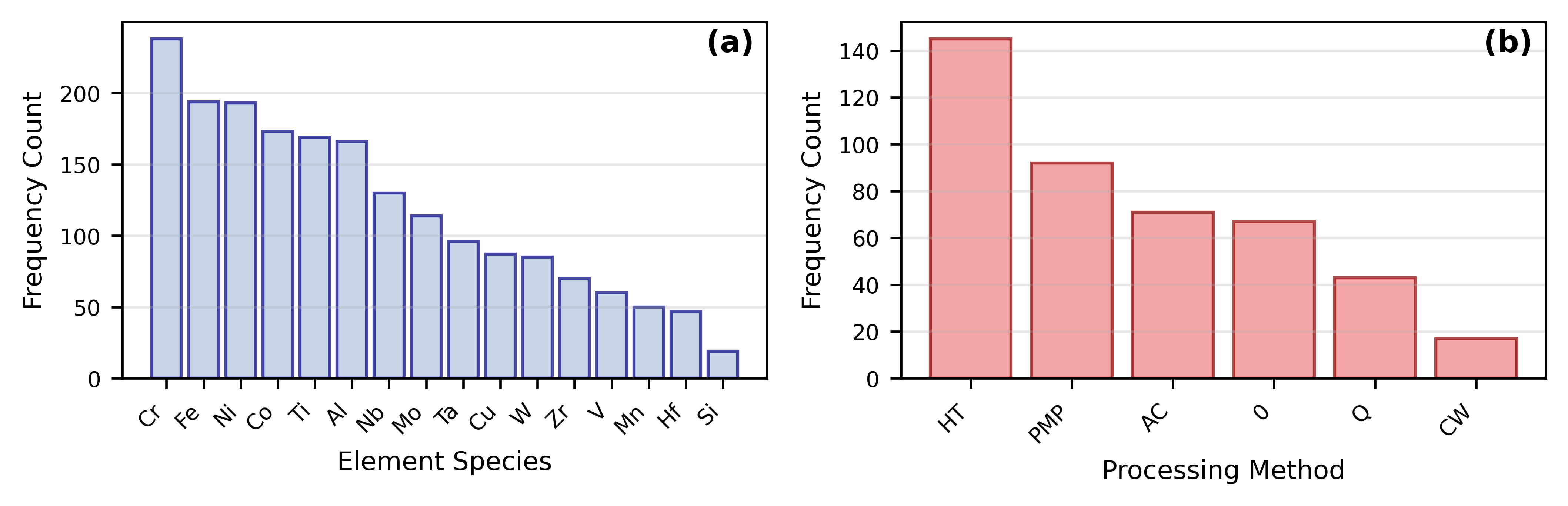}
    \caption{(a) The distribution of elements of alloys in the dataset. (b) The distribution of the processing treatments of alloys in the dataset, including: "HT" (heat-treatment), "PMP" (powder metallurgical processes), "AC" (As-cast alloys without further treatments), "Q" (Quenched) and "0" (Processes that cannot be classified as any other processes)}
    \label{fig:fullwidth}
\end{figure}

To demonstrate the general effectiveness of contextual embeddings as a representation for processing treatment information, three different kinds of models are trained to investigate the effectiveness of embedding descriptor set on improving model performance, which are random forest models with composition and temperature descriptors (RF), models with composition, temperature and one-hot encoded generalized processing symbol descriptors (RF-S) and models with composition, temperature and embedding descriptors (RF-E). The performance of the models under cross fold evaluation is plotted in Fig. 4. The model performance of RF-E and RF-S models are compared against RF model to benchmark the effectiveness of different ways of expression of HEAs' processing treatment information. The RF-S models demonstrate poorer performance compared to the RF model, suggesting that generalized processing symbols not only fail to provide sufficient processing condition information relevant to alloy hardness but also deteriorate model performance by introducing irrelevant features that contribute to overfitting. The RF-E models, however, improves the $\mathrm{R^2}$ by over 0.1 and decreases mean squared error (MSE) by \textasciitilde18\% compared baseline RF model. The failure of RF-symbol model in successfully represent alloys' processing information in highlights the complicated and subtle nature of alloys' processing treatment information that is far beyond the representative capability of one-hot encoding based representation of conventional tabulated datasets. On the other hand, the superior performance of the RF-E models compared to the RF and RF-S models demonstrates the effectiveness of contextual embeddings as a medium of representation capable of capturing the subtle and complicated alloy processing treatment information.

\begin{figure}[H]
    \centering
    \includegraphics[width=\textwidth]{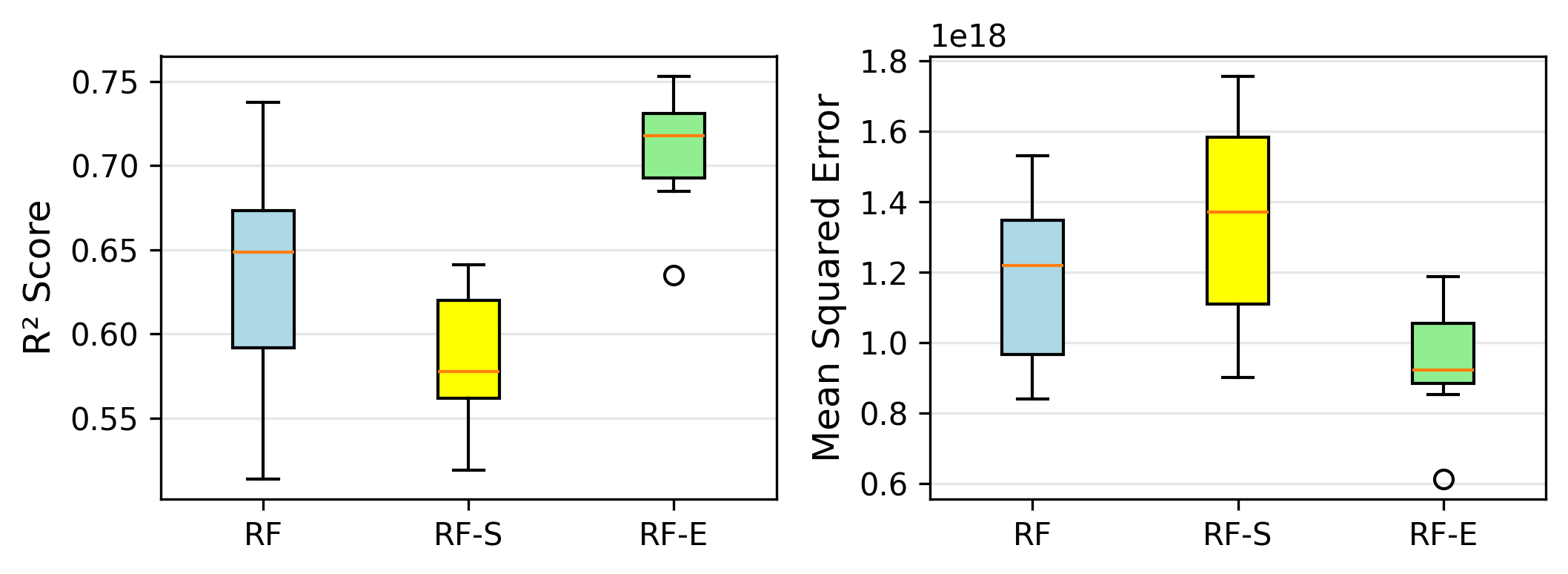}
    \caption{The $\mathrm{R^2}$ and MSE of RF, RF-S and RF-E models}
    \label{fig:fullwidth}
\end{figure}

We then proceed to compare the performance of different classic machine learning regressor models on different datasets, which are raw dataset with only composition and temperature, the dataset with processing symbol added on top of the raw dataset and the dataset with embeddings instead of processing symbol to investigate how the embeddings are utilized by different types of models. The performances of decision tree, elastic net, XGBRegressor, random forest and multilayer perceptron models trained on different types of models are evaluated and plotted in Fig. 5. From the figure it can be observed that elastic net yields much superior performances on the dataset with embeddings (green) than baseline dataset (blue), indicating that the processing-property relationship is largely linear such that linear elastic net models could successfully utilize the embedding descriptors to improve prediction. However, decision tree models and multilayer perceptron models are not successful in utilizing processing treatment information from embeddings. As decision tree models are susceptible to random splitting and MLP models have great amounts of parameters to be fitted, both are susceptible to over fitting. In contrast, both random forest and xgboost regressors can utilize the ensemble of trees and elastic net can utilize the inherent regularization to generalize better. This suggests that processing treatment information is encoded in a low dimensional subspace of the embeddings, such that models with implicit or explicit feature selection — through regularization or ensemble methods — are necessary to extract it effectively.

Further analysis is conducted to see which natural language processing techniques would be the most suitable format of natural language representation of alloys' processing history information. Bag of Words, TF-IDF, FastText and Google gemini embeddings are implemented to represent the natural language description of HEAs' processing history, with the vocabulary of the first two methods being the entirety of the HEA processing treatment corpus minus the english stop words. All four of them demonstrates better performance than baseline models, in which no processing history information is incorporated. Interestingly, even vocabulary-based methods like TF-IDF and bag of words demonstrates performances superior to baseline. This indicates that the processing history information could be utilized by these methods. Besides this, Gemini embedding demonstrates almost no improvement compared to TF-TDF. This suggests that the relevant information is vocabulary-based instead of being semantic-based. Also, it is worth noting that TF-IDF outperforms FastText, which is a word-to-vector embedding model trained on general corpus and represents a paragraph through pooling and encodes general semantic knowledge instead of material-science specific knoledge. This suggests that the models are capable of utilizing domain-specific information of alloys relevant to the actual processing treatment of HEA instead of simply utilizing irrelevant and general information contained in the description.

\begin{figure}[H]
    \centering
    \includegraphics[width=\textwidth]{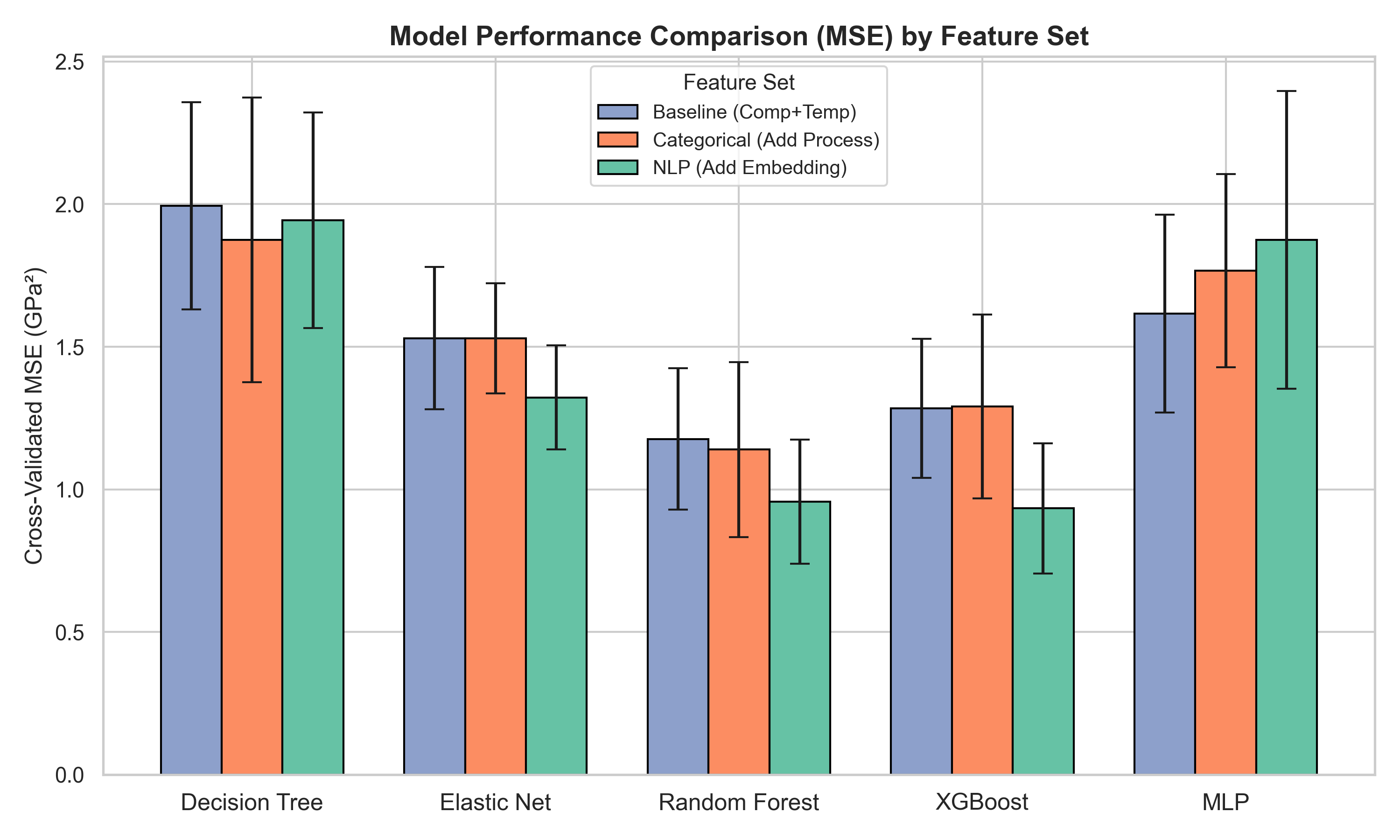}
    \caption{The performance of different classic machine learning models trained on the dataset. The baseline dataset contain compositional and temperature information, the categorecal dataset adds one-hot encoded symbols for processing treatment descriptions, and the NLP dataset adds embedding of alloy processing information description on top of baseline dataset}
    \label{fig:fullwidth}
\end{figure}

\begin{figure}[H]
    \centering
    \includegraphics[width=\textwidth]{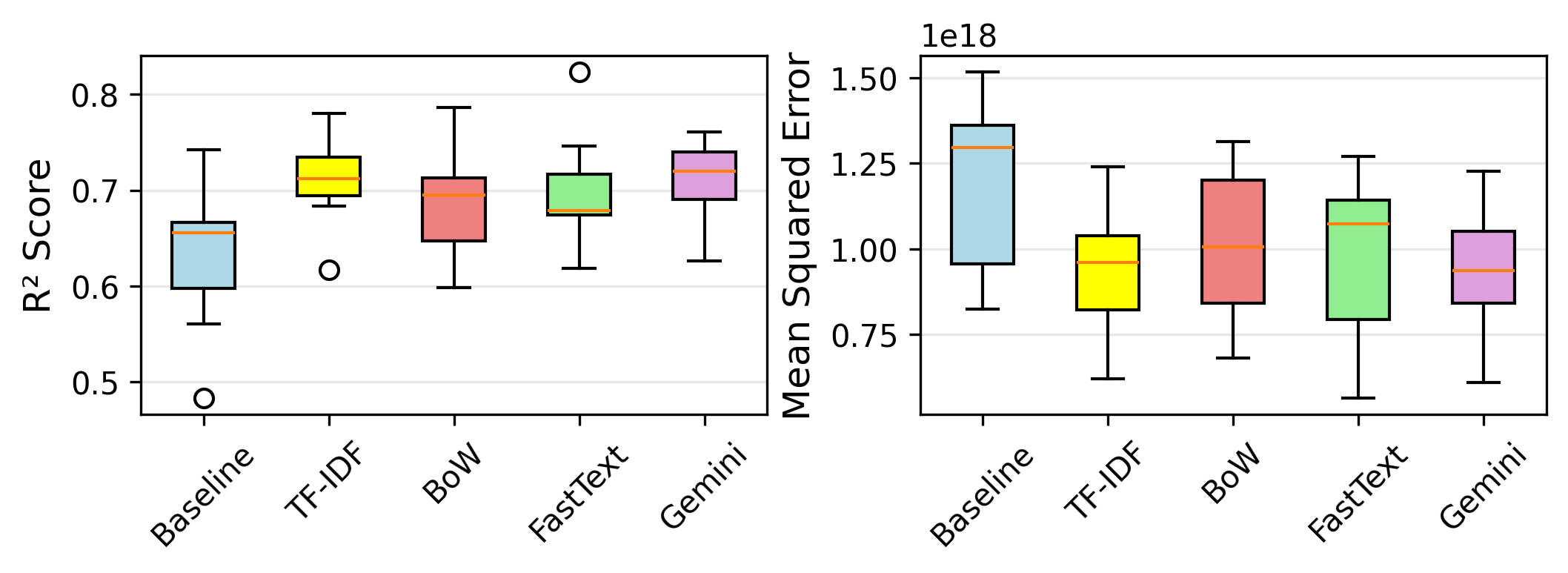}
    \caption{The performance of models trained with HEA processing treatment information represented with different natural language processing techniques}
    \label{fig:fullwidth}
\end{figure}

In conclusion, we applied natural language processing techniques to acquire vector representations of HEA processing treatment. Our modeling demonstrates that these representations are effective at expressing HEA processing treatment information and can be utilized to effectively improve HEA property predictions.

\bibliography{ref}   

\end{document}